\documentclass[12pt]{article}
\usepackage{graphicx}

\newcommand\pubnumber{SNSN-323-63}
\newcommand\pubdate{\today}

\def\Title#1{\begin{center} {\Large #1 } \end{center}}
\def\Author#1{\begin{center}{ \sc #1} \end{center}}
\def\Address#1{\begin{center}{ \it #1} \end{center}}

\newcommand\pubblock{\rightline{\begin{tabular}{l} \pubnumber\\
         \pubdate  \end{tabular}}}
\newenvironment{Abstract}{\begin{quotation}  }{\end{quotation}}
\newenvironment{Presented}{\begin{quotation} \begin{center} 
             PRESENTED AT\end{center}\bigskip 
      \begin{center}\begin{large}}{\end{large}\end{center} \end{quotation}}
\def\Acknowledgements{\bigskip  \bigskip \begin{center} \begin{large}
             \bf ACKNOWLEDGEMENTS \end{large}\end{center}}
\def\beq{\begin{equation}}
\def\eeq#1{\label{#1}\end{equation}}
\def\eeqn{\end{equation}}

\def\beqa{\begin{eqnarray}}
\def\eeqa#1{\label{#1}\end{eqnarray}}
\def\eeqan{\end{eqnarray}}

\let\bar=\overbar

\def\Dslash{\not{\hbox{\kern-4pt $D$}}}
\def\dslash{\not{\hbox{\kern-2pt $\del$}}}

\def\msb{{\bar{\ssstyle M \kern -1pt S}}}

\newcommand{\SU}{{\rm SU}}

\newcommand{\MeV}{{\rm MeV}}

\newcommand{\ignore}[1]{} %VERY USEFUL. NEVER USE WITH verbatim

\begin{document}
\begin{titlepage}
\pubblock

\vfill
\Title{Charm and Strangeness with Heavy-Quark Spin Symmetry}
\vfill
\Author{L.Tolos$^{1,2}$, C. Garcia-Recio$^3$, J. Nieves$^4$, O. Romanets$^5$ and L.L. Salcedo$^3$}
\Address{ $^1$ Instituto de Ciencias del Espacio (IEEC/CSIC), Campus Universitat 
Aut\`onoma de Barcelona, Facultat de Ci\`encies, Torre C5, E-08193 Bellaterra 
(Barcelona), Spain\\
$^2$ Frankfurt Institute for Advanced Studies, Johann Wolfgang Goethe University, Ruth-Moufang-Str. 1,
60438 Frankfurt am Main, Germany \\
$^3$Departamento de F{\'\i}sica At\'omica, Molecular y Nuclear, 
Universidad de Granada, E-18071 Granada, Spain \\
$^4$Instituto de F{\'\i}sica Corpuscular (centro mixto CSIC-UV),
Institutos de Investigaci\'on de Paterna, Aptdo. 22085, 46071, Valencia, Spain \\
$^5$Theory Group, KVI, University of Groningen,
Zernikelaan 25, 9747 AA Groningen, The Netherlands 
}
\vfill
\begin{Abstract}
We study charmed and strange baryon resonances that are generated dynamically within a unitary meson-baryon coupled-channel model which incorporates heavy-quark spin symmetry. This is accomplished by extending the SU(3)
Weinberg-Tomozawa chiral Lagrangian to SU(8) spin-flavor symmetry and implementing a strong flavor symmetry breaking. The model generates dynamically resonances with negative parity in all the isospin, spin, and
strange and charm sectors that one can form from an s-wave interaction between pseudoscalar and vector meson multiplets with $1/2^+$ and $3/2^+$ baryons. Our results are compared with experimental data from several facilities as well as with other theoretical models. Moreover, we obtain the properties of charmed pseudoscalar and vector mesons in dense matter within this coupled-channel unitary effective model by taking into account Pauli-blocking effects and meson self-energies in a self-consistent manner. We obtain the open-charm meson spectral functions in this dense nuclear environment, and discuss their implications on the formation of $D$-mesic nuclei at FAIR energies.
\end{Abstract}
\vfill
\begin{Presented}
Charm2012 \\
The 5th International Symposium on Charm Physics
\end{Presented}
\vfill
\end{titlepage}
\def\thefootnote{\fnsymbol{footnote}}
\setcounter{footnote}{0}

\section{Introduction}

The nature of new charmed and strange hadron resonances is an active topic of research, with data coming from  CLEO, Belle, BaBar  and other experiments \cite{facility00} as well as the planned experiments such as PANDA and CBM at the FAIR facility at GSI \cite{gsi00}. The pursued goal is to understand whether those states can be described with the usual three-quark baryon or quark-antiquark meson interpretation or, alternatively, qualify better as hadron molecules.

Recent approaches based on coupled-channel dynamics have proven to be very successful in describing the existing experimental data. In particular, unitarized coupled-channel methods have been applied in the meson-baryon sector with charm content \cite{Tolos:2004yg,Lutz:2003jw,Hofmann:2005sw,Hofmann:2006qx, Mizutani:2006vq,JimenezTejero:2009vq}, partially motivated by the parallelism between the $\Lambda(1405)$ and the $\Lambda_c(2595)$. Other existing coupled-channel approaches are based on the J\"ulich meson-exchange model \cite{Haidenbauer:2007jq,Haidenbauer:2010ch} or on the hidden gauge formalism \cite{Wu:2010jy}.

Nonetheless, these models are not consistent with heavy-quark spin symmetry (HQSS) ~\cite{Isgur:1989vq,Neubert:1993mb,Manohar:2000dt}, which is a proper QCD symmetry that appears when the quark masses, such as the charm mass, become larger than the typical confinement scale. Aiming to incorporate heavy-quark symmetry, an SU(8) spin-flavor symmetric model has been  recently developed \cite{GarciaRecio:2008dp,Gamermann:2010zz}, which includes vector mesons similarly to the SU(6) approach developed in the light sector of Refs.~\cite{GarciaRecio:2005hy,Toki:2007ab}. The model can generate dynamically resonances with negative parity in all the isospin, spin, strange and charm sectors  that one can form from an s-wave interaction between pseudoscalar and vector meson multiplets with $1/2^+$ and $3/2^+$ baryons \cite{Romanets:2012hm}. In the following we will show some results in the $C=1,S=0$ sector.

We will then focus on the modifications of the dynamically-generated states in the nuclear medium and on the properties of open charm mesons in a dense nuclear environment.  The behaviour of the dynamically-generated resonances in the nuclear medium and the consequences for charmed meson in dense matter will be also part of the  physics program of the PANDA and CBM experiments at the  future FAIR facility at GSI \cite{gsi00}.   In this paper we will show the open-charm meson spectral functions in this nuclear environment within a self-consistent approach in coupled channels, and discuss their implications on the formation of $D$-mesic nuclei  at FAIR energies.

\section{SU(8) WT model with HQSS}
\label{su8wt}

HQSS predicts that all types of spin interactions vanish for infinitely massive quarks: the dynamics is unchanged under arbitrary transformations in the spin of the heavy quark. Thus, HQSS connects vector and pseudoscalar mesons containing charmed quarks. On the other hand, chiral symmetry fixes  the lowest order interaction between Goldstone bosons and other hadrons  in a model independent way; this is the Weinberg-Tomozawa (WT) interaction. Thus, it is appealing to have a predictive model for four flavors including all basic hadrons (pseudoscalar and vector mesons, and $1/2^+$ and $3/2^+$ baryons) which reduces to the WT interaction in the sector where Goldstone bosons are involved and which incorporates heavy-quark spin symmetry in the sector where charm quarks participate. This is a model assumption which is justified in view of the reasonable semiqualitative outcome of the SU(6) extension in the three-flavor sector \cite{Gamermann:2011mq} and on a formal plausibleness on how the SU(4) WT interaction in the charmed pseudoscalar meson-baryon sector comes out in the vector-meson exchange picture.

We use the model extension of the WT SU(3) chiral Lagrangian \cite{Romanets:2012hm}. The model obeys SU(8) spin-flavor symmetry and also HQSS in the sectors where the number of $c$- and $\bar{c}$- quarks are conserved separately. Schematically,
\begin{equation}
\label{symbolic}
{\mathcal L}^{SU (8)}_{ \rm WT} 
=
 \frac{1}{f^2} [[M^{\dagger} \otimes M]_{\bf 63_{a}}
  \otimes [B^{\dagger} \otimes B]_{ \bf 63} ]_{ \bf 1} ,
\end{equation}
which represents a $t$-channel coupling to the ${\bf 63_{a}}$ (antisymmetric) representation of the mesons of the ${\bf 63}$ SU(8) representation and to the $ {\bf 63}$ of baryons falling in the ${\bf 120}$. In the $s$-channel, the meson-baryon space reduces into four SU(8) irreps, from which two multiplets ${\bf 120}$ and ${\bf 168}$ are the most attractive. As a consequence, dynamically-generated baryon resonances are most likely to occur in those sectors, and therefore only states which belong to these two representations will be studied in the following. 

The SU(8)-extended WT meson-baryon interaction reads
\begin{equation}
V_{ij}(s)= D_{ij}
\frac{2\sqrt{s}-M_i-M_j}{4\,f_i f_j} \sqrt{\frac{E_i+M_i}{2M_i}}
\sqrt{\frac{E_j+M_j}{2M_j}} 
\,.
\label{eq:vsu8break}
\end{equation}
Here, $i$ ($j$) are the outgoing (incoming) meson-baryon channels. The quantities $M_i$, $E_i$ and $f_i$ stand for the mass of the baryon, the center of mass energy of the baryon, and the meson decay constant in the $i$ channel, respectively. $D_{ij}$  are the matrix elements coming from the SU(8) group structure of the coupling for the various charm, strange, isospin and spin ($CSIJ$) sectors, which display exact SU(8) invariance. However, this symmetry is severely broken in nature, so we implement a symmetry-breaking mechanism. The symmetry breaking pattern, with regards to flavor, follows the chain $\SU(8)\supset\SU(6)\supset\SU(3)\supset\SU(2)$, where the last group refers to isospin. The symmetry breaking is introduced by means of a deformation of the mass and decay constant parameters.  This allows us to assign well-defined SU(8), SU(6) and SU(3) labels to the resonances and to find HQSS invariant states.

In order to calculate the scattering amplitudes, $T_{ij}$, we solve the on-shell Bethe-Salpeter equation in coupled channels using the interaction matrix $V$ as kernel:
\begin{equation}
\label{LS}
 T(s) =\frac{1}{1-V(s)G(s)} V(s).
\end{equation}
$G(s)$ is a diagonal matrix containing the meson-baryon propagator for each channel.  $D$, $T$, $V$, and $G$ are matrices in coupled-channel space. The loop function $G(s)$ is logarithmically ultraviolet divergent and it is regularized by the means of subtraction point regularization \cite{Nieves:2001wt}.

The dynamically-generated baryon resonances can be obtained as poles of the scattering amplitudes. Close to the pole, the $T$-matrix behaves as
\begin{equation}
\label{Tfit}
 T_{ij} (s) \approx \frac{g_i g_j}{\sqrt{s}-\sqrt{s_R}}
\,,
\end{equation}
with the mass ($m_R$) and the width ($\Gamma_R$) given by $\sqrt{s_R}=m_R - \rm{i} \Gamma_R/2$, while $g_i$ is the coupling to the different meson-baryon channels.

\section{Dynamically-generated baryon states}
\label{dyn}

Our model predicts dynamically generated states in different charm and strange sectors  \cite{GarciaRecio:2008dp,Gamermann:2010zz,Romanets:2012hm}. We have assigned to some of them a tentative identification with known states from the PDG \cite{Nakamura:2010zzi}. This identification is made by comparing the data from the PDG on these states with the mass, width and, most important, the coupling to the meson-baryon channels of our dynamically-generated poles. In the following we comment on the $C=1,~S=0$ sector.
 
In this work we obtain the three lowest-lying states of Ref.~\cite{GarciaRecio:2008dp} in the $\Lambda_c$ sector, which come from the most attractive SU(8) representations. However, those states appear with slightly different masses due  to the different subtraction point, and the use of slightly different $D_s$ and $D_s^*$ meson decay  constants. The experimental $\Lambda_c(2595)$ resonance can be identified with the ${\bf  21}_{2,1}$ pole that we found around $2618.8\,\MeV$, as similarly done in Ref.~\cite{GarciaRecio:2008dp}. The width in our case is, however, bigger because we have not changed the subtraction point to fit its position as done in Ref.~\cite{GarciaRecio:2008dp}.  In any case,  we have not included the three-body decay channel $\Lambda_c \pi \pi$ \cite{Nakamura:2010zzi}. We also observe a second broad $\Lambda_c$ resonance at 2617~MeV with a large coupling to the open channel $\Sigma_c \pi$, very close to $\Lambda_c(2595)$. This is precisely the same two-pole pattern found in the charmless $I = 0, S = -1$ sector for the $\Lambda(1405)$\cite{Jido:2003cb}. The third spin-$1/2$ $\Lambda_c$ resonance is found around 2828~MeV  and cannot be assigned to any experimentally known resonance. We also find one spin-$3/2$ resonance in this sector located at $(2666.6 -i 26.7\,\MeV)$. Similarly to Ref.~\cite{GarciaRecio:2008dp}, this resonance is assigned to the experimental $\Lambda_c(2625)$. A similar resonance was found at $2660\,\MeV$ in the $t$-channel vector-exchange model of Ref.~\cite{Hofmann:2006qx}. The novelty of our calculations is that we obtain a non-negligible contribution from the vector meson-baryon channels to the generation of this resonance.

Three $\Sigma_c$ resonances are obtained for $C=1, S=0, I=1, J=1/2$ with masses 2571.5, 2622.7 and 2643.4~MeV and widths 0.8, 188.0 and 87.0~MeV, respectively,  which nicely agree with the three lowest lying resonances found in Ref.~\cite{GarciaRecio:2008dp}. These are predictions of our model since there is no experimental confirmation yet. The model of Ref.~\cite{JimenezTejero:2009vq} predicts the existence of two resonances in this sector. However, only one of them can be identified to one of ours but with a strong vector meson-baryon component.
We predict two spin-$3/2$ $\Sigma_c$ resonances. The first one, a bound state at 2568.4~MeV, is thought to be the charmed counterpart of the $\Sigma(1670)$. The second state at $2692.9 -i 33.5$ ~MeV has not a direct experimental comparison.

\section{Charmed mesons in dense nuclear matter}
\label{medium}

In this section we analyze the behavior of the dynamically-generated states in the $(C=1,~S=0)$ sector within the nuclear medium. In particular, we study the properties of open charm ($D$ and $D^*$) mesons in a dense nuclear environment related to the modification of dynamically-generated  $\Lambda_c$ and $\Sigma_c$ states in this environment.

The self-energy and, hence, spectral function for $D$ and $D^*$ mesons are obtained self-consistently in a simultaneous manner, as it follows from HQSS by taking, as bare interaction, the SU(8)-extended WT described in Sec.~\ref{su8wt}. The in-medium solution incorporates Pauli blocking effects and open charm meson self-energies in the intermediate propagators \cite{tolos09}. 

%%%%%%%%%%%%%%%%%%%%%%%%%%%%%%%%
\begin{figure}
\begin{center}
\includegraphics[width=0.43\textwidth,height=6.5cm]{art_spec.eps}
\hfill
\includegraphics[width=0.43\textwidth, angle=90]{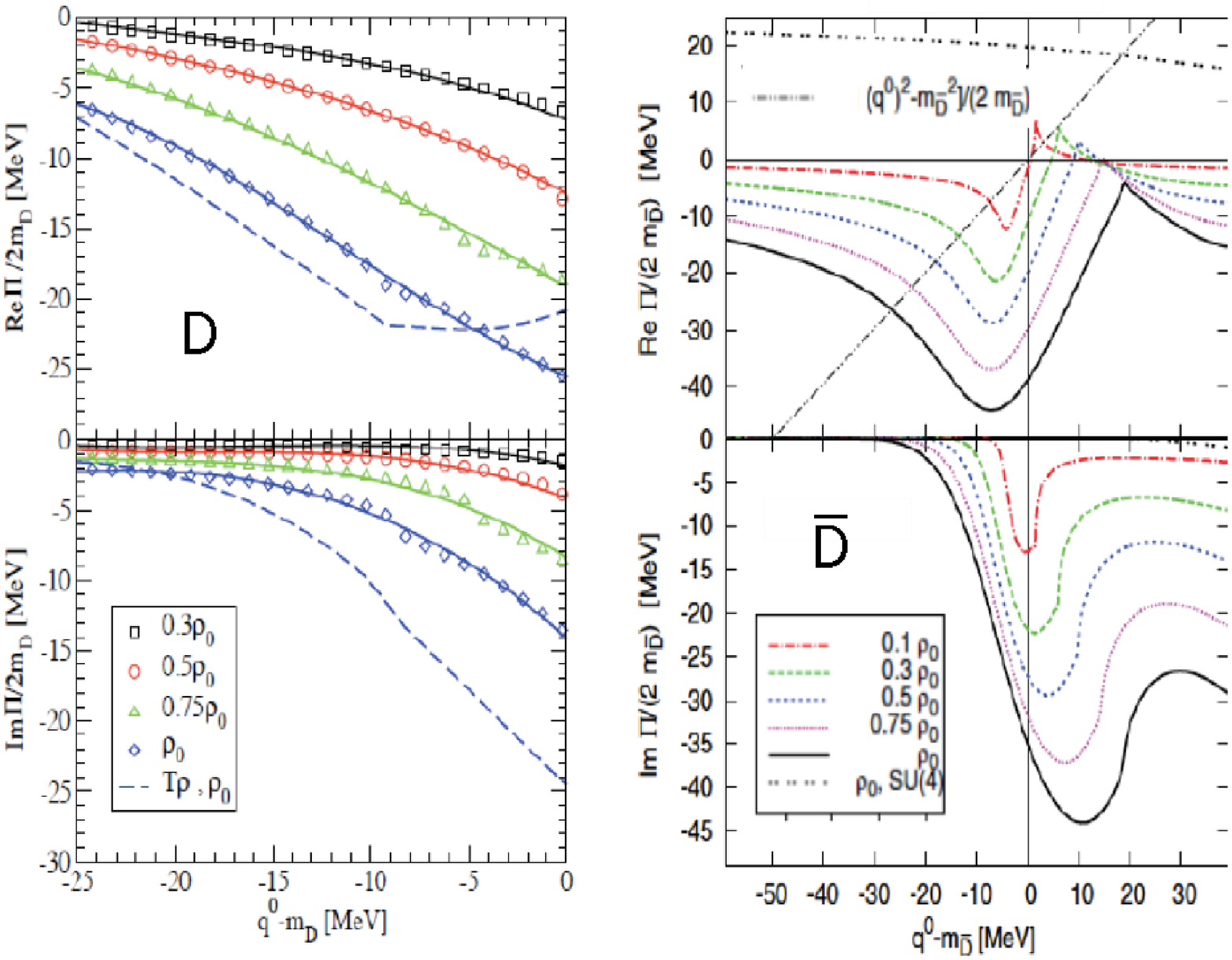}
\caption{Left: $D$ and $D^*$ spectral functions in nuclear matter at $\vec{q}=0$ MeV/c. Right: $D$ and $\bar D$ optical potential at $\vec{q}=0$ MeV/c for different densities }
\label{fig1}
\end{center}
\end{figure}

%%%%%%%%%%%%%%%%%%%%%%%%%%%%%%%%%%%%%

The $D$ and $D^*$ self-energies are obtained summing the transition amplitude for the different isospins over the nucleonic Fermi sea, $n(\vec{p})$: 
\begin{eqnarray}
\Pi_D(q_0,{\vec q}) &=& \int \frac{d^3p}{(2\pi)^3}\, n(\vec{p})  \, [\, {T_{DN}}^{(I=0,J=1/2)} +
3 \, {T_{DN}}^{(I=1,J=1/2)}\, ]  \ ,  \nonumber \\ 
\Pi_{D^*}(q_0,\vec{q}\,) &=& \int \frac{d^3p}{(2\pi)^3} \, n(\vec{p}\,) \,
\, \left [~ \frac{1}{3} \, {T}^{(I=0,J=1/2)}_{D^*N}+
{T}^{(I=1,J=1/2)}_{D^*N}+ \right . \nonumber \\
&&  \left . \frac{2}{3} \,
{T}^{(I=0,J=3/2)}_{D^*N}+ 2 \,
{T}^{(I=1,J=3/2)}_{D^*N}\right ] \  ,
\label{eq:selfd}
\end{eqnarray}
\noindent
where the $T$-matrices are evaluated at $s=P_0^2-{\vec P}^2$, being  $P_0=q_0+E_N(\vec{p})$ and $\vec{P}=\vec{q}+\vec{p}$ are the total energy and momentum of the meson-nucleon pair in the nuclear matter rest frame, and ($q_0$,$\vec{q}\,$) and ($E_N$,$\vec{p}$\,) stand  for the energy and momentum of the meson and nucleon, respectively, in this frame. The self-energy is determined self-consistently since it is obtained from the
in-medium amplitude which contains the meson-baryon loop function, and this quantity itself is a function of the self-energy. Then, the meson spectral function  reads
\begin{eqnarray}
S_{D(D^*)}(q_0,{\vec q})= -\frac{1}{\pi}\frac{{\rm Im}\, \Pi_{D(D^*)}(q_0,\vec{q})}{\mid
q_0^2-\vec{q}\,^2-m_{D(D^*)}^2- \Pi_{D(D^*)}(q_0,\vec{q}) \mid^2 } \ .
\label{eq:spec}
\end{eqnarray}

%%%%%%%%%%%%%%%%%%%%%%%%%%%%%%%
\begin{figure}
\begin{center}
\includegraphics[width=0.34\textwidth,angle=-90]{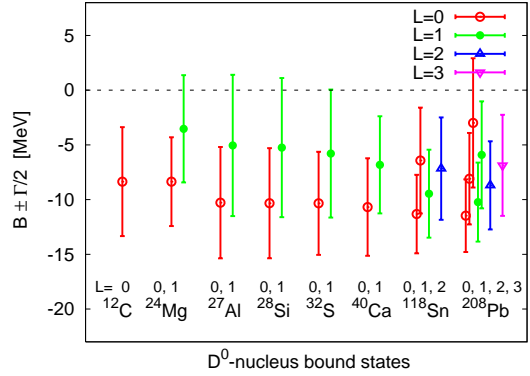}
\caption{$D^0$-nucleus bound states with defined angular momentum $L$. \label{fig2}}
\end{center}
\end{figure}
%%%%%%%%%%%%%%
On the l.h.s of  Fig.~\ref{fig1} we display the $D$ and $D^*$ spectral functions, which show a rich spectrum of resonance-hole states. The $D$ meson quasiparticle peak mixes strongly with $\Sigma_c(2823)N^{-1}$
and $\Sigma_c(2868)N^{-1}$ states while the $\Lambda_c(2595)N^{-1}$ is clearly visible in the low-energy tail. The $D^*$ spectral function incorporates the $J=3/2$ resonances, and the quasiparticle peak fully mixes with the  $\Sigma_c(2902)N^{-1}$ and $\Lambda_c(2941)N^{-1}$ states.  As density increases, these $Y_cN^{-1}$ modes tend to smear out and the
spectral functions broaden with increasing phase space, as seen before in the $SU(4)$ model \cite{Mizutani:2006vq}.
Note that not all the states described in Sec.~\ref{dyn} are seen in the $D$ and $D^*$ spectral functions, as some of them do not strongly coupled to $DN$ and/or $D^*N$ states. Moreover, as we have just seen, resonances with higher masses than those described in Sec.~\ref{dyn} are also present in the spectral functions. Those resonant states were obtained from the wider energy range explored in  Ref.~\cite{GarciaRecio:2008dp}.

%%%%%%%%%%%%%%%%%%%%
\begin{figure}
\begin{center}
\includegraphics[width=0.47\textwidth]{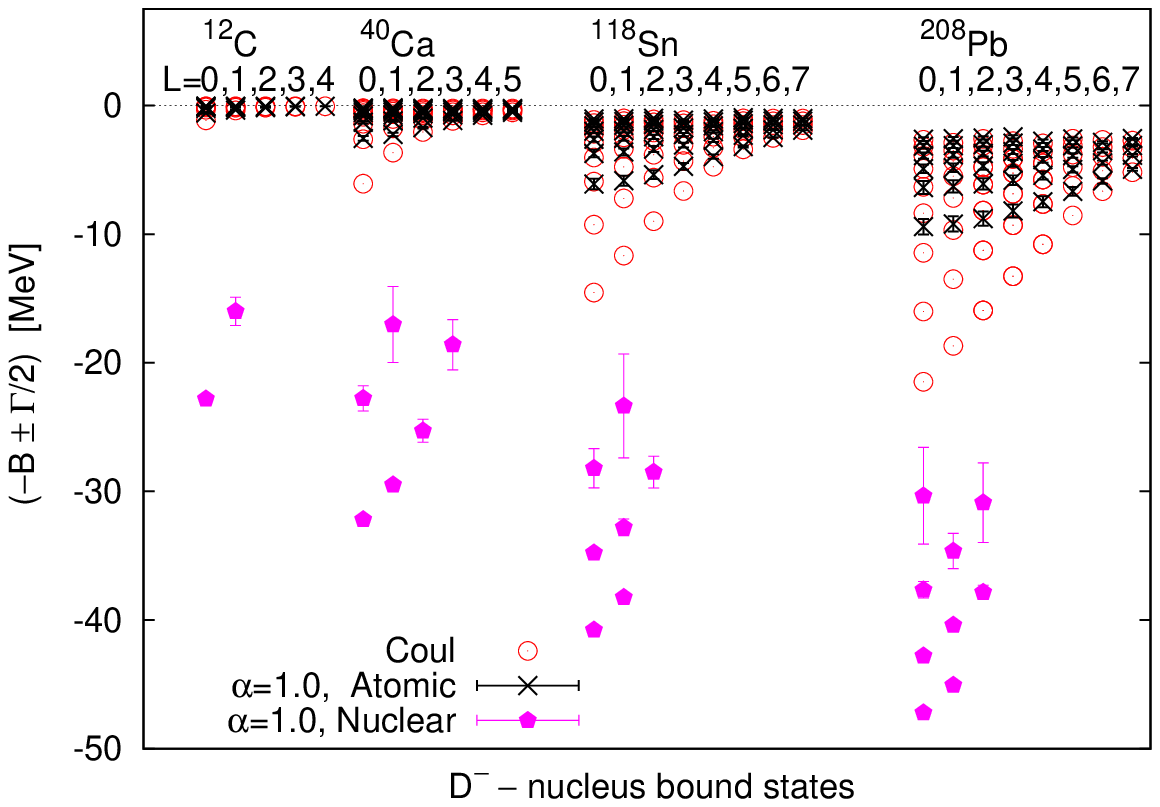}
\hfill
\includegraphics[width=0.47\textwidth]{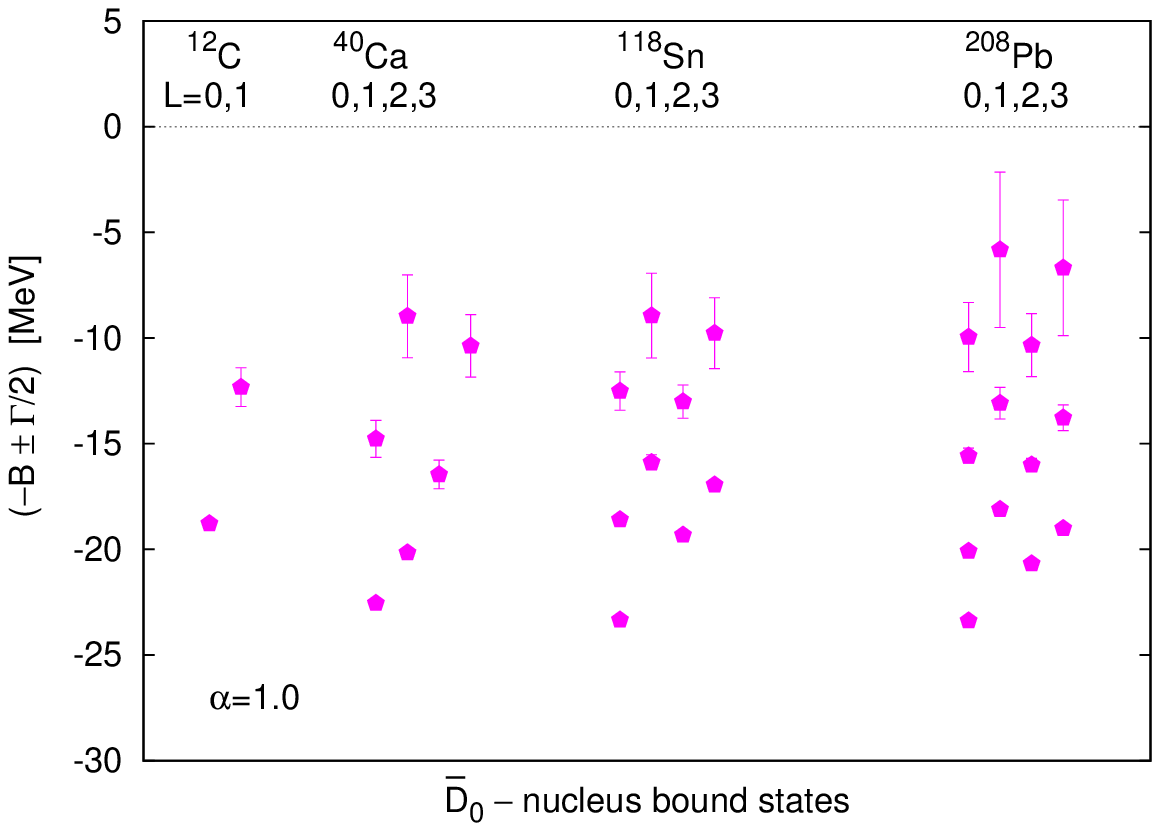}
\caption{$D^-$ and $\bar D^0$- nucleus bound states with defined angular momentum $L$. \label{fig3}}
\end{center}
\end{figure}

%%%%%%%%%%%%%%%%%%%%%%%%%%%%%%%%%%%%%5

\section{D-mesic nuclei}

The possible formation of $D$ and $\bar D$-meson bound states in $^{208}$Pb was predicted in Ref.~\cite{tsushima99} relying upon an attractive  $D$ and $\bar D$ -meson potential in the nuclear medium  based on a quark-meson coupling (QMC) model \cite{sibirtsev99}. The experimental observation of those bound states, though, might be problematic since, even if there are bound states, their widths could be very large compared to the separation of the levels. This is indeed the case for the potential derived from a $SU(4)$ $t$-vector meson exchange model for $D$-mesons \cite{TOL07}.

In order to analyze the possible formation of bound states with charmed mesons in nucleus, we solve the Schr\"odinger equation in the local density approximation. For that purpose, we take the energy dependent optical potential 
\begin{equation}
  V(r,E) = \frac{
  \Pi(q^0=m+E,\vec{q}=0,~\rho(r))}{2 m},
\label{eq:UdepE}
\end{equation}
where $E=q^0-m$ is the $D$ or $\bar D$ energy excluding its mass, and $\Pi$ the meson self-energy. The optical potential for different densities is shown on the r.h.s of Fig.~\ref{fig1}. For $D$ mesons we observe a strong energy dependence of the potential close to the $D$ meson mass due to the mixing of the quasiparticle peak with the  $\Sigma_c(2823)N^{-1}$ and $\Sigma_c(2868)N^{-1}$ states. As for the $\bar D$ meson, the presence of a bound state at 2805 MeV \cite{Gamermann:2010zz}, almost at $\bar D N$ threshold, makes the potential also strongly energy dependent in contrast to the SU(4) model.

The question is whether $D$ and/or $\bar D$  will be bound in nuclei. We start by analyzing $D$ mesons in nuclei.  In Fig.~\ref{fig2} we show $D^0$-meson bound states in different nuclei. We observe that the $D^0$-nucleus states are weakly bound, in contrast to previous results using the QMC model, and have significant widths \cite{carmen10}, in particular, for $^{208}$Pb \cite{tsushima99}. Only $D^0$-nucleus bound states are possible since the Coulomb interaction prevents the formation of observable bound states for $D^+$ mesons.

With regard to $\bar D$-mesic nuclei, we observe in Fig.~\ref{fig3}  that not only $D^-$ but also $\bar{D}^0$ bind in nuclei. The spectrum 
contains states of atomic and of nuclear types for all nuclei. The nuclear states exist for lower angular momenta only. Compared to the pure Coulomb levels, the atomic states are less bound and the nuclear ones are more bound and may present a sizable width \cite{GarciaRecio:2011xt}.

%%%%%%%%%%%%%%%%%%%%%%%%%%%%%%%
\begin{figure}
\begin{center}
\includegraphics[width=0.43\textwidth]{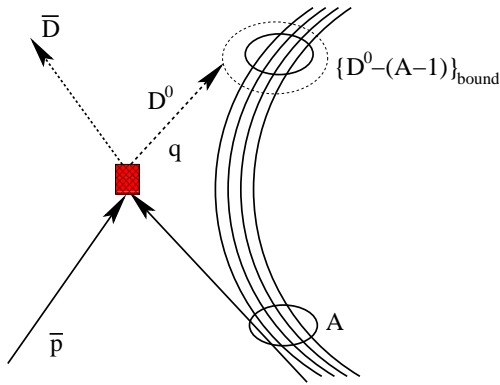}
\caption{Possible mechanism for production of
  $D^0$-mesic nuclei with an antiproton beam\label{fig4}}
  \end{center}
\end{figure}

%%%%%%%%%%%%%%%%%%%%%%%%%%%%%%%%%%%%%

The information on bound states is very valuable to gain some knowledge on the charmed meson-nucleus interaction, which is of interest for PANDA at FAIR. The experimental detection of $D$ and $\bar D$-meson bound states is, though, a difficult task. For example, reactions with antiprotons on nuclei for obtaining $D^0$-nucleus states (see Fig.~\ref{fig4}) might have a very low production rate. Similar reactions but with proton beams, although difficult, seem more likely to trap a $D^0$ in nuclei \cite{carmen10}.

\section{Conclusions and Outlook}

We have studied charmed and strange baryon resonances that are generated dynamically within a unitary meson-baryon coupled-channel model which incorporates heavy-quark spin symmetry.  Our results have been compared with experimental data from several facilities,  as well as with other theoretical models. Moreover, the properties of open-charm mesons in dense matter have been obtained. The in-medium solution  accounts for Pauli blocking effects and meson self-energies. We have analyzed the evolution with density of the open-charm meson spectral functions. We have finally studied the possible formation of $D$-mesic nuclei. On one hand, only  weakly bound $D^0$-nucleus states seem to be feasible. On the other hand, $D^-$ and $\bar D^0$- nuclear bound states are possible, the latter ones with a stronger binding than for $D^0$ and with also a sizable width. The experimental detection is, though, most likely a challenging task.

\Acknowledgements
This research was supported by DGI and FEDER
funds, under Contract Nos. FIS2011-28853-C02-02,
 FIS2011-24149, FPA2010-16963 and the
Spanish Consolider-Ingenio 2010 Programme CPAN
(CSD2007-00042), by Junta de Andalucõa Grant
No. FQM-225, by Generalitat Valenciana under Contract
No. PROMETEO/2009/0090 and by the EU
HadronPhysics2 project, Grant Agreement No. 227431.
O. R.  wishes to acknowledge support from the
Rosalind Franklin Programme. L. T. acknowledges support
from Ramon y Cajal Research Programme, and from FP7-
PEOPLE-2011-CIG under Contract No. PCIG09-GA-
2011-291679.

\end{document}